\documentclass[final,5p,times,twocolumn]{elsarticle}

\usepackage{graphicx}

\usepackage{amssymb}

\biboptions{}

\journal{Solid State Communications}

\begin{document}

\begin{frontmatter}


\title{Title\tnoteref{label1}}
\author[label1]{Rayda Gammag\corref{cor1}}
\ead{rayda.gammag@apctp.org}
\author[label2]{Cristine Villagonzalo}
\cortext[cor1]{Corresponding author}
\address[label1]{Asia Pacific Center for Theoretical Physics, POSTECH,
San 31 Hyoja-dong, Nam-gu, Pohang, Gyeongbuk 790-784, Korea}
\address[label2]{Structure and Dynamics Group, National Institute of Physics, University of the Philippines, 
Diliman, 1101 Quezon City, Philippines}

\title{Two-dimensional electron gas tilt-induced Landau level crossings}




\begin{abstract}
This work elucidates the novel behavior found in a two-dimensional electron
gas (2DEG) under a tilted magnetic field in which the field's angle becomes
the dominant factor in tuning the spin-splitting rather than the strength
of the spin-orbit interaction. The 2DEG eigenvalues are derived with Rashba
and Zeeman interactions for various tilt angles and they show crossing-free
levels except at very high tilt. Moreover, concomitant with the crossings
is the appearance of beats in the 2DEG density of states. The crossings from
different levels occur consecutively at around $87^{\circ}$.  Similar new
observations in Shubnikov-de Haas experimental measurements by
Hatke \textit{et al.} \cite{Hatke2012} attributed such phenomena to an
in-plane-magnetic-field-induced increase in the effective mass.
We show here that this behavior is inherent to a 2DEG where spin-orbit
interaction and the in-plane magnetic field contribution are taken into account.
\end{abstract}


\end{frontmatter}



\section{Introduction}
The advent of spintronics \cite{Datta1990} and the quest to control spin in 
semiconductors have fueled the current research studies 
on Landau and spin energy levels of a two-dimensional electron gas (2DEG)
\cite{Becker2010,Xia2011,SSC2012,EPJB2012,Hatke2012}. 
Such investigations offer insight on how these energy levels and the spin-orbit 
coupling of electrons are affected by an applied magnetic field and the 2DEG 
plane orientation. Tilting of the 2DEG plane relative to the field direction 
was considered to enhance spin-orbit effects \cite{EPJB2012,Desrat2005} making 
them easier  to detect \cite{Lipparini2006}.

Early magnetotransport studies in Ga$_x$In$_{1-x}$As/InP \cite{Koch1993} at 
millikelvin temperatures in strong magnetic fields at a tilt angle $\theta$ of the order of $\sim 80^{\circ}$ 
showed coincidence of the first two Landau levels (LLs) with low carrier concentration, 
while the overlap disappears at sufficiently high mobility. This was believed to be caused 
by electron-electron interactions which shift the system from a spin-unpolarized state to a spin-polarized 
state \cite{Koch1993}.
Other novel effects of tilting have since been observed which are not yet completely understood. 
%
%
For instance, the fractional quantum Hall states in the 2DEG in GaAs/AlGaAs heterostructures
were found to be suppressed as the sample's tilt angle $\theta$ is increased,
and resurging only at $\theta=76^{\circ}$ with a possible 2DEG nematic phase \cite{Xia2011}. 
Moreover, experiments on the de Haas-van Alphen (dHvA) effect in high mobility 2DEG 
yield magnetization oscillations exhibiting beating patterns only at large tilt angles ($\theta \geq 75^{\circ}$) 
\cite{Wilde2009}. The authors attributed these features to spin splitting induced 
by the structure inversion asymmetric potential in the heterostructure.
This type of coupling contribution is known as the Rashba spin-orbit interaction 
(SOI) \cite{Rashba1960}. Usually the Rashba SOI is more dominant than the 
bulk-inversion-asymmetry in III-V \cite{Giglberger2007}, II-VI \cite{Gui2004} and 
Si-based \cite{Wilamowski2002} semiconductors. This offers a favorable opportunity 
for practical applications because the Rashba SOI can be controlled via a gate voltage.

In a different study carried out on a 2DEG near the surface of a highly doped 
\textit{p}-type InSb crystal, measurements on the differential conductivity ($dI/dV$) also 
manifested beating patterns of the LL intensity for different 
magnetic fields \cite{Becker2010}. Their results on the local density of states
were consistent with the Rashba spin splitting.
Not only is the Rashba SOI linked with the beats in these Shubnikov-de Haas (SdH) oscillations
 but it also gives way to a direction-dependent energy spectrum \cite{Entin2008}. 
The authors of this present paper have previously shown that  when both Rashba and Zeeman 
interactions are taken into account there are no beating patterns 
in the density of states of the 2DEG for $\theta\leq 80^{\circ}$ \cite{SSC2012}. 
The authors observed that depending on the intensity of 
the applied magnetic field $B$, the Rashba SOI and Zeeman term signatures 
can be distinguished \cite{EPJB2012}. Moreover, at the tilt angles below
$80^{\circ}$, the specific heat exhibited a suppressed beating pattern 
as a function of $B$.  It was not expected then that large tilt angles could give way to
more novel observations.

In this current work, we will distinguish the effects of $\theta$ from the rest of 
the tunable parameters. 
To the best of our knowledge, this is the first work that clearly
identifies the effects of the direction of $\vec{B}$.  
We will show how  the energy spectrum of a two-dimensional 
electron gas changes as we sweep through the different directions of $\vec{B}$.
Special attention is devoted to the large tilt case, that is when
$\theta\rightarrow 90^{\circ}$, as it reveals the beats that seem to have 
only been restrained in the smaller angle regime \cite{SSC2012,EPJB2012}.  
Another interesting feature considered is that of a \textit{crossing} 
which occurs when two adjacent LLs become degenerate 
for a given $\vec{B}$, Rashba SOI strength, and tilt angle. 
We will show that these crossings are correlated to the beating patterns 
observed in thermodynamic quantities \cite{Wilde2009,Luo1990}. 

Recently SdH oscillations of ultrahigh mobility 
2DEG in GaAs/AlGaAs quantum wells showed beating patterns at high tilt angles 
($\geq 87^{\circ}$) which  indicate consecutive level crossings occurring at 
the same angle \cite{Hatke2012}. They explained their results as due to an increase 
in the carrier mass induced by the in-plane field with crossings dependent 
on the filling factor \cite{Hatke2012}.
Here we will show that the density of states of a 2DEG model which incorporates 
Rashba SOI naturally exhibits multi-crossings at large tilt angles.
From this a causal relationship of the LL crossings with 
the beating patterns observed in thermodynamic quantities is corroborated.

\section{Crossing States}
\begin{figure}
\vspace{0.6cm}
\centering
{\includegraphics[width=3.2in]{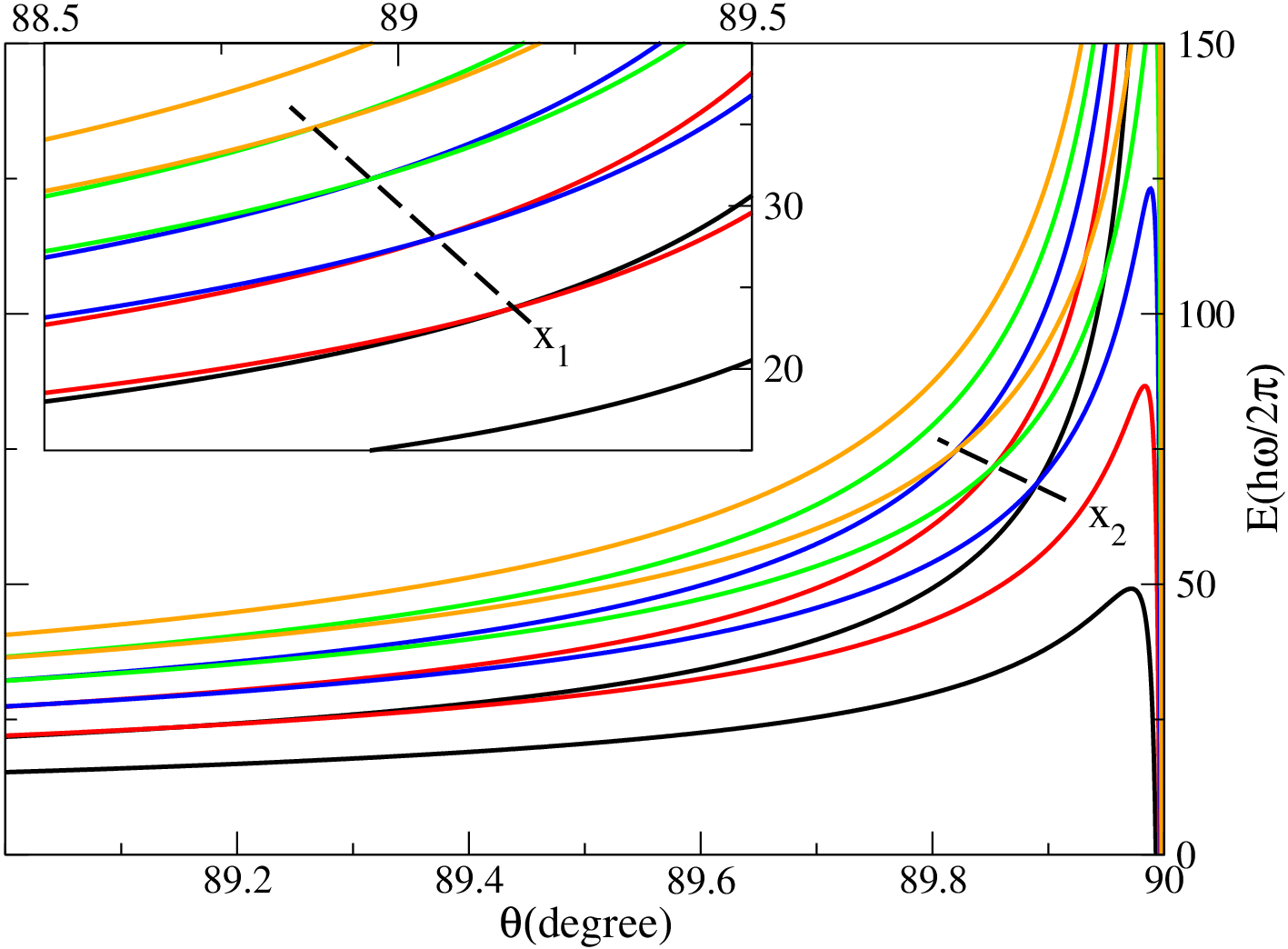}}
\caption{The eigenvalues as a function of the tilt angle for $n \in [1,5]$.  
A pair of identically colored lines indicates the two spin states of a given 
LL $n$. The dashed line in the inset points to the first crossing while
that at the main frame the second.  
Here $\alpha = 3.5 \times 10^{-11}$ eVm and $B = 0.1$ T.}
\label{fig:Etheta}
\end{figure}

If Zeeman effect is the lone source of splitting, the SdH or dHvA effect is 
characterized by a single frequency \cite{Das1989,Studenikin2005}.  
This is because, just like the Zeeman term, the cyclotron energy $\hbar \omega_c$ 
is also linearly proportional with the magnetic field magnitude $B$ \cite{Das1989,Winkler2003}, 
that is, $\omega_c = eB_{\perp}/m^*$. The perpendicular component $B_{\perp}$ 
is set along the $z-$axis and is equal to $B_z = B\cos(\theta)$ and $m^{*}$ is 
the effective mass. When the Rashba SOI is significant, the spin splitting is no longer 
simply linear with $B$. The Hamiltonian describing a single electron in this system 
can be written as
\begin{equation}
H = H_0 + H_R + H_Z = \frac{\hbar^2 \vec{k}^2}{2m^*} +  
\alpha (\vec{\sigma} \times \vec{k}) \cdot \hat{z} - \vec{\mu} \cdot \vec{B},
\label{H}
\end{equation}
where $H_0$ is the free particle energy, $H_R$ is the Rashba SOI, and $H_Z$ is the Zeeman energy.  
Hence, on top of the strongly degenerate LLs, additional degeneracies can arise
if the 2DEG further experiences a spin-orbit interaction.
The strength of the Rashba SOI is indicated by the parameter $\alpha$ which is assumed 
to be constant in this work. 
Here $\vec{\sigma}$ are the Pauli matrices and the magnitude of the wave 
vector $\vec{k}$ is determined by $k_j = -i \nabla_j + \frac{e}{ \hbar} A_j$, 
where $A_j$ is the $j$-th component of the vector potential. The Hamiltonian 
above is exactly solvable when $\vec{B}$ is normal to the 2DEG. 
However, if a finite in-plane component, $B_{\parallel} = B\sin(\theta)$, is present, 
the problem becomes underspecified and an analytic solution  can only be 
achieved when additional specifications are imposed \cite{SSC2012,Bychkov1990}. 

\begin{figure}
\vspace{0.6cm}
\centering
{\includegraphics[width=3.2in]{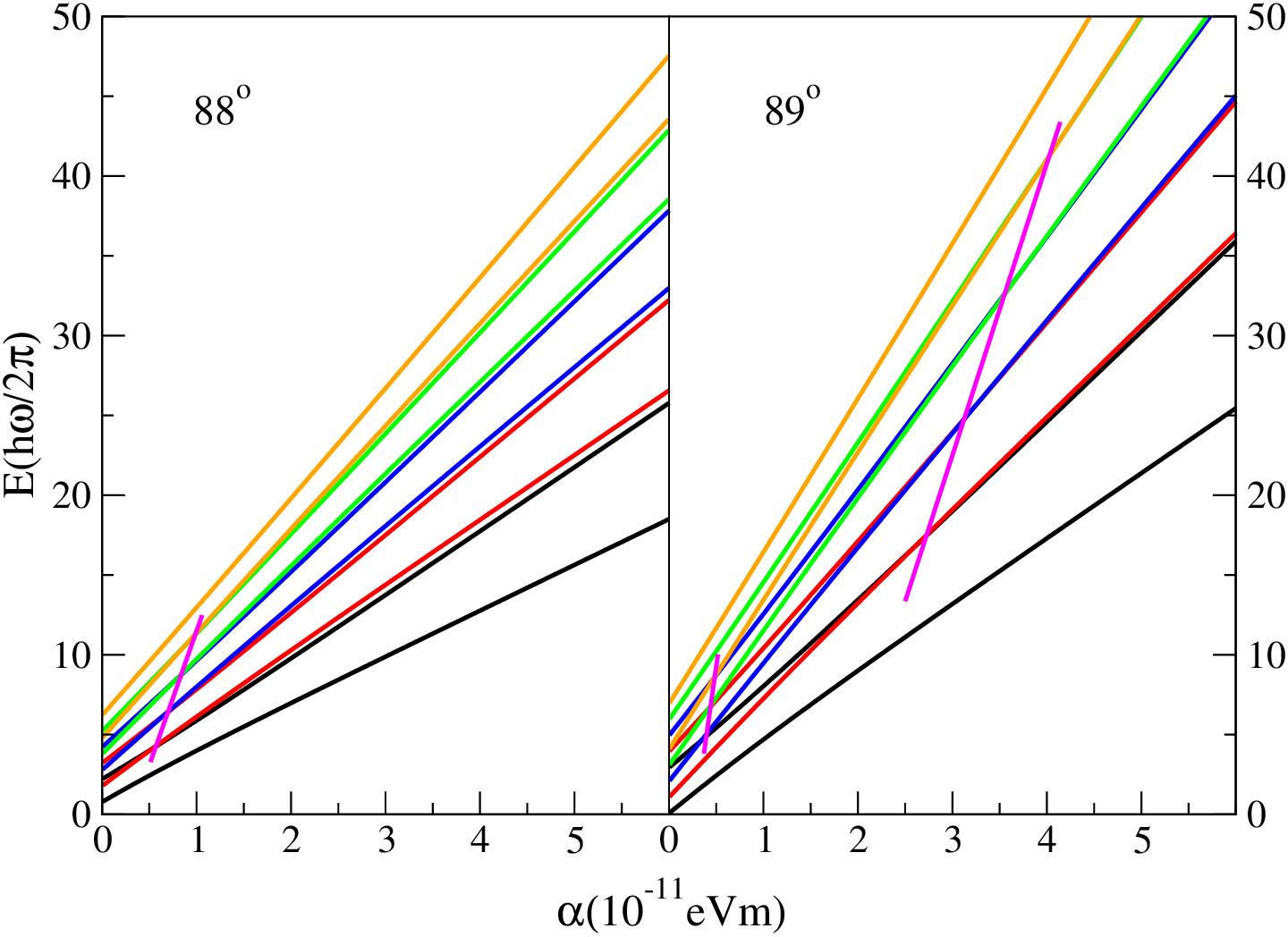}}
\caption{The eigenvalues as a function of the Rashba SOI for $n \in [1,5]$.  
Here $B = 0.1$ T. A pair of identically colored lines indicate the two spin 
states of a given LL. The magenta lines serve as guides to the eye pointing to the crossing points.}
\label{fig:Ealpha2}
\end{figure}

We circumvent the underspecified problem by imposing equal probability among pairs
of LLs that are known to cross when $B_{\parallel} = 0$ \cite{SSC2012,EPJB2012}.
This yields an exact expression for the eigenvalues which is given as
\begin{eqnarray}
E_n^{\pm} =  n + \frac{1}{2} + \frac{\Upsilon}{2} \left(\sqrt{n+1} + \sqrt{n} \right) \nonumber\\
\pm \frac{1}{2} \sqrt{\left[\Upsilon(\sqrt{n + 1} - \sqrt{n}) + 2 \Omega_z \right]^2 
+ 4 \Omega_+ \Omega_-}.
\label{Eq:Evalues}
\end{eqnarray}
Here $n$ is the LL index, $\Omega_j = \mu_B B_j/\hbar \omega_c$ is 
the $j$-th component of the rescaled Zeeman energy, 
$\Upsilon = 2 \alpha \sqrt{\zeta / \hbar \omega_c}$ 
is the rescaled Rashba parameter, $\zeta = m^*/2 \hbar^2$ 
and $\Omega_{\pm} = \Omega_x \pm i \Omega_y$.  
In the results that follow, energy units are given in terms of the cyclotron energy 
$\hbar \omega_c$. 
In addition, the 2DEG is considered to be on the $x-y$ plane subject to an applied
field in the $x-z$ direction with magnitude $B = \sqrt{B_x^2 + B_z^2}$. In general,
$m^{*}$ is dependent on the energy due to nonparabolicity but in this work
we let $m^{*}=0.05\times m_e$ ($m_e$ is the free electron mass) which is the
effective mass in a 2DEG in InGaAs heterostructures \cite{Nitta2007} and is of the
same order as other semiconductor systems.  

In some theoretical studies \cite{Bychkov1990,Jiang2009}, the degeneracies at 
the crossing levels are lifted as $\theta$ deviates from zero. 
This in fact is the case here.  However, when the large tilt regime is approached
crossings begin to reappear as illustrated in fig.~\ref{fig:Etheta}.
Although not included in the figure's scope, a crossing-free 
spectrum is observed in the region $\theta < \theta_{x_1}$ where $\theta_{x_1}$ is the 
angle where the first crossing $x_1$ shows up. As $\theta$ is further increased 
more crossings begin to emerge.  The crossing point $x$ shifts to lower $\theta$ 
as the LL index $n$ is increased. This is expected since higher $n$ 
corresponds to a wider spin splitting  as indicated in eq.~(\ref{Eq:Evalues}). 

Inspecting eq.~(\ref{Eq:Evalues}), we anticipate an increasing spin gap $\Delta$ 
with increasing $\theta$. This gap between the two spin levels of a given LL 
($E_n^+ - E_n^-$) expands until they cross their adjacent LL.  
The crossings, however, occur only at very large $\theta$ as shown in the figure.  

It has been predicted that if a finite $B_{\parallel}$ is present, 
the energy levels do not cross as the Rashba SOI parameter is increased 
\cite{Bychkov1990}. We check the range of validity of this prediction 
regarding the energy levels as a function of the Rashba parameter, $E(\alpha)$, 
using our result in Eq.~(\ref{Eq:Evalues}). Indeed, for $\theta < \theta_{x_1}$, 
$E(\alpha)$ is crossing-free \cite{SSC2012}.  However, for $\theta \geq \theta_{x_1}$, 
$E(\alpha)$ starts to exhibit crossings.  This is depicted in Fig.~\ref{fig:Ealpha2}. 
At $88^{\circ}$, only one crossing is resolved because of the slow gap widening.
Relatively larger angles, $89^{\circ}$ for example, exhibit faster spread 
causing the occurence of multi-crossings. From the plot we understand 
that the tilt angle has a suppressing effect when $\theta < \theta_{x1}$ but it gives way to 
crossings once $\theta \geq \theta_{x1}$.

Comparing Fig.~\ref{fig:Etheta} and Fig.~\ref{fig:Ealpha2}, we can deduce that large $\theta$
and small $\alpha$ both favor crossings. Increasing the Rashba parameter widens both the spin
gap and the Landau gap.  The interplay of these two produces the crossings at small
$\alpha$ and the absence of crossings at large $\alpha$ as depicted in Fig.~\ref{fig:Ealpha2}.
Whether Rashba SOI gives way or inhibits crossings is determined by $\theta$.
Note that in both figures, the Zeeman contribution is a constant for a given $\theta$.

\begin{figure}
\vspace{0.7cm}
\centering
{\includegraphics[width=3.25in]{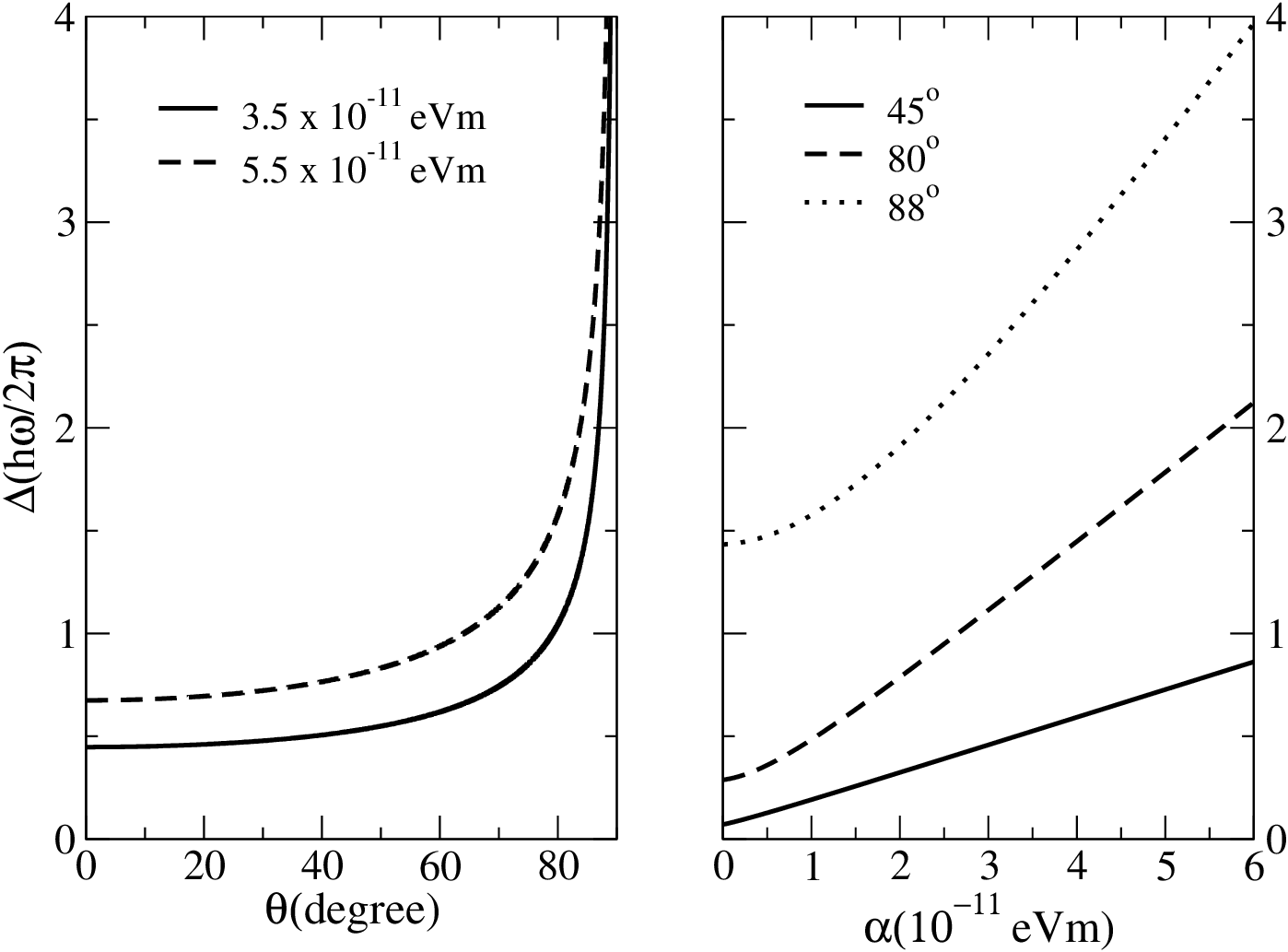}}
\caption{The spin splitting as a function of the tilt
(left plot) and the Rashba parameter (right plot) when $B = 0.1$ T.}
\label{fig:delta}
\end{figure}

Figure \ref{fig:delta} exhibits in detail how the spin gap varies with the Rashba
SOI and the field's tilt angle. For the left panel, the
$\Delta(\theta)$ curves for two values of $\alpha$ are qualitatively similar in shape.
Only the magnitude is increased upon increasing $\alpha$. However, at
very large angles ($\theta > 85^{\circ}$) the two curves approach the same set of values.
This signifies that, although the Rashba SOI strength brings about spin splitting of adjacent levels, 
the large field tilting tunes the splitting behavior in addition to $\alpha$. 
For the right panel, on the other hand, $\Delta(\alpha)$ is linear with respect to $\alpha$ 
at small angles.
We have observed that as $\theta$ increases $\Delta(\alpha)$ gradually shifts
from a linear to a quadratic behavior with $\alpha$ .  Further increasing $\theta$, 
specifically at $\theta = 88^{\circ}$, $\Delta$ has a cubic dependence on $\alpha$ in addition to
its increased magnitude. Figure \ref{fig:delta} affirms how the primary control
belongs to $\theta$.

\section{Density of States}
In experiments, the density of states (DOS) are usually probed indirectly through 
measurements of transport properties such as  the magnetization \cite{Wilde2009}
and the specific heat \cite{Gornik1985}. Recent developments have allowed the
localized DOS to be probed in real space through scanning tunneling spectroscopy
\cite{Becker2010,Hashimoto2008}. The 2DEG spatially averaged spectrum obtained 
in Ref.~\cite{Becker2010} has a beating pattern of the LL intensity 
in the first subband which the authors attributed to the Rashba spin
splitting. Motivated by this observation, together with the suppressed
beating patterns in the specific heat at small and intermediate angles
\cite{EPJB2012}, and the enhanced beats in the magnetization at large
tilts \cite{Wilde2009}, the perceptible influence of $\theta$
on the DOS in the presence of the Rashba SOI will be considered and shown here.
This is done using the  phenomenological form of DOS as defined by a 
series of Gaussian functions, namely,
\begin{equation}
\mbox{DOS}(E) = \frac{eB_{\perp}}{h} \sum_n \left(\frac{1}{2\pi} \right)^{1/2} 
\frac{1}{\Gamma} \, \exp \left[-\frac{(E - E_n^{\pm})^2}{2 \Gamma^2} \right],
\label{Eq:DOS} 
\end{equation}
where $\theta$ affects the eigenvalues $E_n^{\pm}$ and the magnitude of $B_{\perp}$.

The broadening width $\Gamma$ measures how the DOS deviates from its 
ideal impurity-free 2DEG behavior. Thus, this parameter is indicative of
the number of impurities and/or the scale of disorder in the system. 
\begin{figure}
\vspace{0.8cm}
\centering
{\includegraphics[width=3.4in]{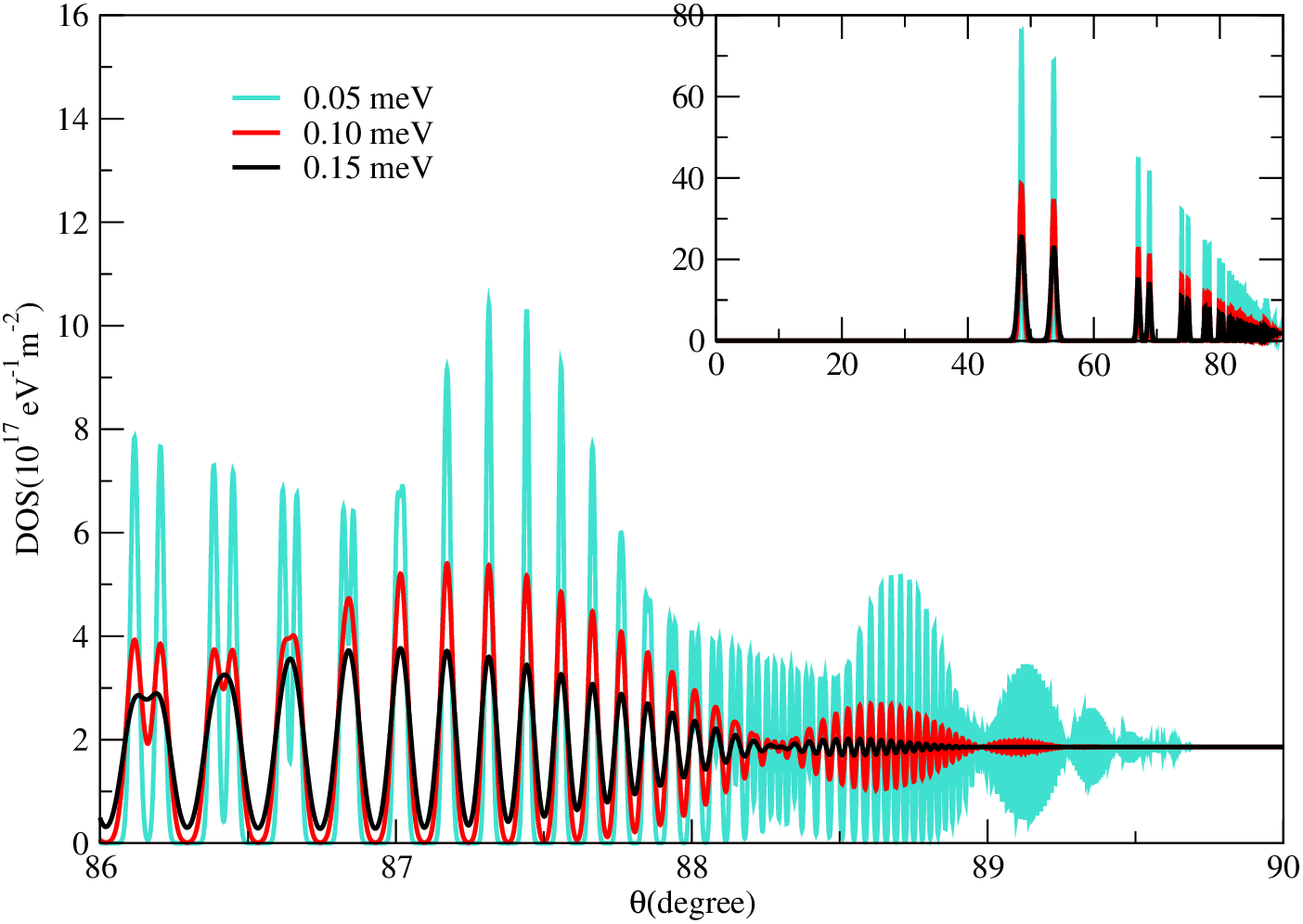}}
\caption{DOS as a function of the tilt angle at $E \approx 16$ meV for 
$\alpha = 3.5 \times 10^{-11}$ eVm, $B = 6$ T, and different 
broadening widths.}
\label{fig:Dtheta}
\end{figure}
Extracted values of the broadening range from 0.4 - 1.5 meV in InAs/GaSb 
\cite{Luo1990}, 1.5 - 3.0 meV in InSb/InAlSb \cite{Gilbertson2009}
and can reach as high as $\sim$20 meV in InSb \cite{Becker2010}
depending on the acceptor concentration. The $\Gamma$ values used here are 
of the same order as the theoretical fits in AlGaAs/GaAs which are
dependent on the field magnitude, 0.1 meV $\times \sqrt{B(\mbox{T})}$. For
simplicity, we keep $\Gamma$ constant. The case of $\Gamma < 1.0$  meV
ensures that the spin-splitting and the tilt effects are not washed out 
and dominated by the broadening.

As shown in the inset of Fig.~\ref{fig:Dtheta}, the DOS peaks monotonously decreased 
for $\theta < 86^{\circ}$. This behavior can be traced to the direct proportionality
of the prefactor of Eq.~(\ref{Eq:DOS}) to $B_{\perp}$. When $\theta$ is large as in
the case of $\theta > 86^{\circ}$, the vanishing amplitude at the beat node 
suggests that the two spin subbands oscillate with similar frequencies.  
As $\theta$ is decreased, the difference 
between the frequencies increases. When $\theta < 86^{\circ}$, for example, 
the two frequencies become very different that only one of them dominates.  
This gives rise to the single-frequency-DOS.  
In Ref. \cite{Luo1990}, the single-frequency oscillation was observed
in the high $B$ region of the spectrum while the two-frequency beating pattern was
detected in the low $B$ region. This is analogous to our observation 
where the latter corresponds to the high $\theta$ domain.  Furthermore, 
we observe that the DOS oscillations with double frequencies, 
namely the beats, become more pronounced as $B$ is intensified. 
This is due to the proportional increase in DOS amplitude. 

Referring still to Fig.~\ref{fig:Dtheta}, we now relate the beats with the crossings.  
Both beats and crossings are absent until $\theta$
reaches a certain large value $( > 86^{\circ})$.  We deduce that 
each DOS antinode is concomitant with a crossing point. 
Since two energy levels merge to the same energy value $E$, the degeneracy of that 
$E$ is doubled. This translates to a maximum DOS envelope function. 
The crossings occur when $\Delta = j\hbar \omega_c$, where $j = 1,2,3, . . .$ 
On the other hand, nodes result when the LLs are 
equidistant to their adjacent levels.  Such a condition is satisfied when 
$\Delta = (j/2) \hbar \omega_c$, where $j$ here corresponds to the order of the node 
\cite{Das1989}.  The latter resembles the usual evenly spaced harmonic oscillator spectrum.

In Fig.~\ref{fig:Dtheta}, we can observe that a large $\Gamma$ can mask the
presence of the beats.  This might help explain the absence of SdH beats in 
materials that are known to have strong Rashba SOI \cite{Brosig2000}.  
In addition, when the tilt angle has not yet reached 
the threshold value $\theta_{x1}$ where the beats begin to emerge, a finite 
$B_{\parallel}$ only suppresses the beats and the SdH is seen to oscillate 
with a monotonously decreasing amplitude.  In this case, for all $\theta < \theta_{x1}$ 
the beats are absent.


Collating all of the abovementioned observations, 
beats are found at the two ends of the tilt spectrum - the zero tilt ($B_{\parallel} = 0$)
\cite{SSC2012} and the large tilt ($B_{\parallel} \rightarrow B$).
In between these two extremities, the beats are not observed. 
Moreover, only at these regions did we also observe the crossings
which is indicative of beats and crossings being concomitant.
Since the beats are due to the spin splitting induced by the Rashba SOI and 
crossings are found in the case of zero tilt \cite{SSC2012,Jiang2009} but are not found for
$\theta < 80^{\circ}$ \cite{SSC2012}, we deduce that the magnetic field's tilt relative to the 2DEG plane
shifts the spin splitting-induced crossings to large angles as observed in this paper.

\begin{figure}
\vspace{0.8cm}
\centering
{\includegraphics[width=3.3in]{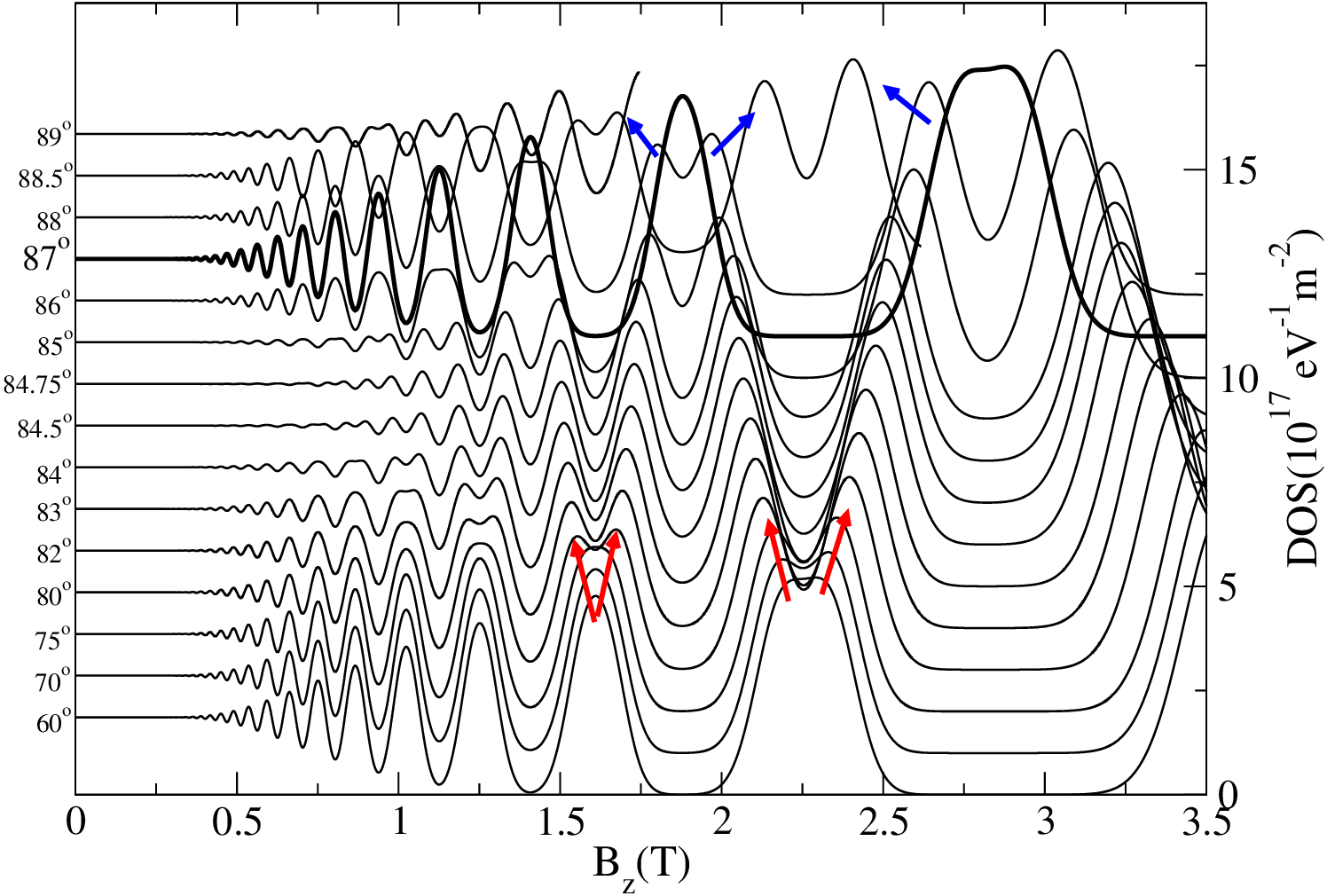}}
\caption{DOS versus $B_{\perp}$ at $E \approx$ 16 meV showing the phase shift, 
the spin splitting and the crossings. The curves are shifted for clarity.  
Here $\alpha = 3.5 \times 10^{-11}$ eVm and $\Gamma = 0.5$ meV.
The red arrows point to the spin splitting of a LL while the 
blue arrows to the reemergence of the spin levels after crossing.}
\label{fig:thetac}
\end{figure}


To quantify how large the large-tilt region is, we plotted DOS verus $B_{\perp} = B_z$.  
See Fig.~\ref{fig:thetac}. We found out that the DOS versus 
$B_{\perp}$ has characteristic 180$^{\circ}$ phase reversals at 
$\theta_{r1} \approx 84.75^{\circ}, \theta_{r2} \approx 88.5^{\circ}$ and $\theta_{r3} \approx 89^{\circ}$.  
The DOS are in-phase for $0 < \theta < \theta_{r1}$, for $\theta_{r1} < \theta < \theta_{r2}$, 
and for  $\theta_{r2} < \theta < \theta_{r3}$.   The complete phase reversal, with the same frequency, however, 
is limited only to the small $B_{\perp}$ region ($B_{\perp} < 1$ T) where the LL splitting is negligible. 
This has been observed experimentally \cite{Tang2009}.
When the splitting becomes evident, it is not straightforward to compare phases because 
the LLs begin to split into two spin levels (See red arrows.).  As $\theta$ is increased, 
the gap between the spin levels widens until they cross other LLs 
(See thickened line at $\theta = 87^{\circ}$.).  
Compared to $\theta = 86^{\circ}$, each amplitude at $\theta = 87^{\circ}$ 
is approximately doubled and the widths of the peaks are narrower. These are
evidences of the crossing of two spin levels of neighboring LLs.  
Note that here we observe multi-level crossings at $\theta = 87^{\circ}$, specifically
at $B_{\perp}=$ 1.126 T, 1.408 T and 1.879 T.
This is in agreement with the experimental data of Hatke \textit{et al.}, in which
this unusual observation of level crossings occurring at the same angle 
are attributed to an in-plane magnetic field-induced increase of the carrier mass.
Indeed in Ref.~\cite{Oto2001} the distortion on a Fermi loop
was directly measured using the magnetic electron 
focusing effect in a 2DEG. This, they say, affects the effective mass
because of the antisymmetry of the cyclotron orbit. 
Although, we kept $m^{*}$ in our work a constant for the simulated data
presented herein, the eigenvalues in Eq.~(\ref{Eq:Evalues}) are derived
from first principles. The exact derivation incorporates the Rashba SOI and 
the in-plane field whose mutual influence on the electron's orbit
changes the characteristics of the energy spectrum and consequently, 
the density of states. 



It is important to mention that we do not observe here the anticrossings 
of adjacent spin-split LLs seen in Ref.~\cite{Desrat2005}. 
The authors of the said paper attributed the anticrossing to the spin-orbit coupling of the collective spin configuration of the 
quantum hall ferromagnets and to the nonparabolicity causing spin mixing.
The authors of \cite{Brosig2000} believe that it does not come from electron-electron 
interactions but from the pronounced nonparobolicity of the bands.
From our work, we infer that anticrossings must come from nonparabolicity 
of the energy bands that are induced by other interactions which are not from the structure 
inversion asymmetric potential and the tilting of the applied field.

In conclusion, we find that the spin gap shifts from a linear
to a polynomial dependence on the powers of $\alpha$ as $\theta$
increases and approaches $90^{\circ}$. Moreover, at these large tilting
we find how Landau level crossings reemerge after being suppressed at lower angles.
These crossings bring about beats in the density of states which are manifestations of 
two spin subbands oscillating at similar frequencies.  In agreement
with recent experimental data from Hatke \textit{et al.}, the crossings are 
found to occur consecutively at around $87^{\circ}$. While they attribute
this unusual behavior to the in-plane magnetic field-induced carrier mass enhancement, 
this work has shown that the multiple crossings are intrinsic to a two-dimensional
electron gas with Rashba spin-orbit interaction and under intense in-plane
magnetic field.





\bibliographystyle{elsarticle-num}



\end{document}